\documentclass[a4paper,11pt]{article}
\pdfoutput=1 

\usepackage{jinstpub}
\usepackage{amsmath,amsfonts}
\usepackage{amssymb}
\usepackage{algorithmic}
\usepackage{graphicx}
\usepackage{textcomp}
\usepackage{xcolor}
\usepackage{subcaption}
\graphicspath{ {pictures/} }
\usepackage{lineno}

\newcommand{\mum}{$\mu m$}

\title{First application of machine learning algorithms to the position reconstruction in Resistive Silicon Detectors}

\author[a,b,1]{F.Siviero,\note{Corresponding author.}}
\author[b,c]{R. Arcidiacono,}
\author[b]{N. Cartiglia,}
\author[a,b]{M. Costa,}
\author[b,c]{M. Ferrero,}
\author[b]{F. Legger,}
\author[b]{M. Mandurrino,}
\author[b]{V. Sola,}
\author[b]{A. Staiano,}
\author[a,b]{M. Tornago,}

\affiliation[a]{Universit\'a degli Studi di Torino, Torino, Italy}
\affiliation[b]{INFN Sezione di Torino, Torino, Italy}
\affiliation[c]{Universit\'a del Piemonte Orientale, Vercelli, Italy}

\emailAdd{federico.siviero@edu.unito.it}

\abstract{
RSDs (Resistive AC-Coupled Silicon Detectors) are n-in-p silicon sensors based on the LGAD (Low-Gain Avalanche Diode) technology, featuring a continuous gain layer over the whole sensor area. The truly innovative feature of these sensors is that the signal induced by an ionising particle is seen on several pixels, allowing the use of reconstruction techniques that combine the information from many read-out channels. 
In this contribution, the first application of a machine learning technique to RSD devices is presented. The spatial resolution of this technique is compared to that obtained with the standard RSD reconstruction methods that use analytical descriptions of the signal sharing mechanism.  A Multi-Output regressor algorithm, trained with a combination of simulated and real data,  leads to a spatial resolution of less than  2 $\mu m$ for a sensor with a 100 $\mu m$ pixel. 
The prospects of future improvements are also discussed.

}


\arxivnumber{2011.02410v1} 

\begin{document}
\maketitle
\flushbottom

\section{Introduction}
RSDs (Resistive AC-Coupled Silicon Detectors) are n-in-p silicon sensors based on the LGAD (Low-Gain Avalanche Diode) technology, featuring a continuous gain layer over the whole sensor active area ~\cite{LGAD}.

The RSD design is an evolution of the LGAD design, as it combines internal gain with internal signal sharing. 
The presence of internal gain assures  the excellent timing performances  of standard LGADs while internal signal sharing enables the ability of precisely reconstruct the impact position  using the information provided by the shared signals. 

Moreover, the RSD design ensures a 100\% fill-factor, meaning that the ratio between the sensor active and total areas is equal to 1, whereas standard LGADs reach a fill-factor ranging from 70\% to 90\% ~\cite{interpad}.

\begin{figure}[h]
\begin{center}
\includegraphics[width=0.7\linewidth]{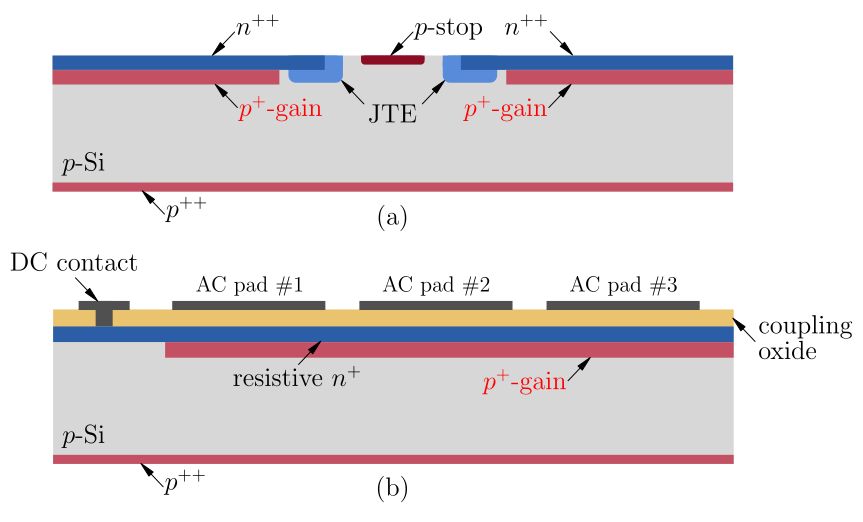}     
\caption{Cross-section of an LGAD (a) compared to an RSD (b). }
\label{fig:LGAD_vs_RSD}
\end{center}
\end{figure}

A comparison between the RSD and LGAD designs is shown in Figure~\ref{fig:LGAD_vs_RSD}.

Signal formation in RSD is based on the principle of AC-coupled resistive read-out. In this design,  the n-doped read-out layer is manufactured to be resistive, i.e., its doping level has been decreased with respect to the standard design. When the signal induced by an ionizing particle on the $n^+$ layer discharges to ground, it uses the lowest impedance path. In resistive read-out, this path is via the AC-pads, grounded by the frontend electronics. This mechanism naturally induces signals on multiple AC pads. The AC read-out is obtained through metal pads  capacitively coupled to the detector's bulk via a dielectric layer deposited between the metal pads and  resistive $n^{+}$ layer. The dielectric layer thickness and the $n^{+}$ resistivity are the parameters governing the signal induction on the pads ~\cite{MM2}.

The sensors measured in this work were part of the first RSD production (RSD1) manufactured at Fondazione Bruno Kessler (FBK). RSD1 features pixelated sensors with different geometries: in this contribution, 3x3 pixel matrices are considered, all with the same $n^{+}$ layer resistivity and dielectric thickness. 4 different geometries are explored: 100-200 \mum, 50-100 \mum, 150-300 \mum, 200-500 \mum, where the first number indicates the pad size and the second one the sensor pitch ~\cite{MM1}.
The difference between the sensor pitch and the pad size will be named \textit{interpad} in the following. As an example, the \textit{interpad} of the 200-500 \mum~ geometry is 300 \mum.

\section{Signal sharing}

Signal formation in RSD is different from that of a standard LGAD. The signal, formed on the $n^+$ resistive layer due to the drift of e/h carriers in the bulk, spreads toward ground. The AC metal pads, grounded via the read-out electronics, offer to fast signals a path to ground with an impedance far lower than that provided by the $n^+$ resistive sheet. For this reason, the fast signal (about 1 ns for 50-$\mu m$ thick sensor) becomes visible on the AC pads, charging the capacitors formed by the AC metal pads and the $n^{+}$ layer, with a delay that increases with distance from the impact point. If the metal pad is right above the impact point, there is no delay. The main advantage of this read-out scheme is that the signal has the same characteristics of a standard LGAD, thus allowing precise timing, and it is shared among multiple pads, which is the key element for precise position reconstruction. 

When a particle crosses the detector, nearby pads see a signal with an amplitude and delay depending on the hit position: closer pads will see a larger and less delayed signal than the further ones. Both delay and attenuation are caused by the propagation of the signal in the resistive layer. 
Signal sharing in RSD is such that there is only one position on the detector surface that produces a given  signal split among 3 or more pads. Therefore 3 pads is the minimum number of active pads required by the reconstruction program to unambiguously identify the particle hit position. As a consequence, all regions within the sensor active area where less than 3 pads are active are inefficient, meaning that the hit position cannot be uniquely identified by the reconstruction program.

Two analytical models of signal sharing in RSD are presented in ~\cite{marta}. The first model, the so-called RSD Main Formula (MF), assumes that the impedance from the particle impact point to the read-out metal pad increases logarithmically with distance. The fraction of the total signal seen by each read-out pad as a function of its distance can be expressed by the following expression:

\begin{equation}
        S_{i}(\alpha_{i},r_{i}) = \frac{\frac{\alpha_{i}}{ln(r_{i})}}{\sum_{j=1}^{n}\frac{\alpha_{j}}{ln(r_{j})}}  
\label{eq:master}
\end{equation}

where $S_{i}$ is the fraction seen by the i-th pad; $\alpha_{i}$ is the angle of view of the pad from the hit position; $r_{i}$ is the distance between the hit position and the metal pad;  $n$ the number of active pads.

\begin{figure}[h]
\begin{center}
\includegraphics[width=0.85\linewidth]{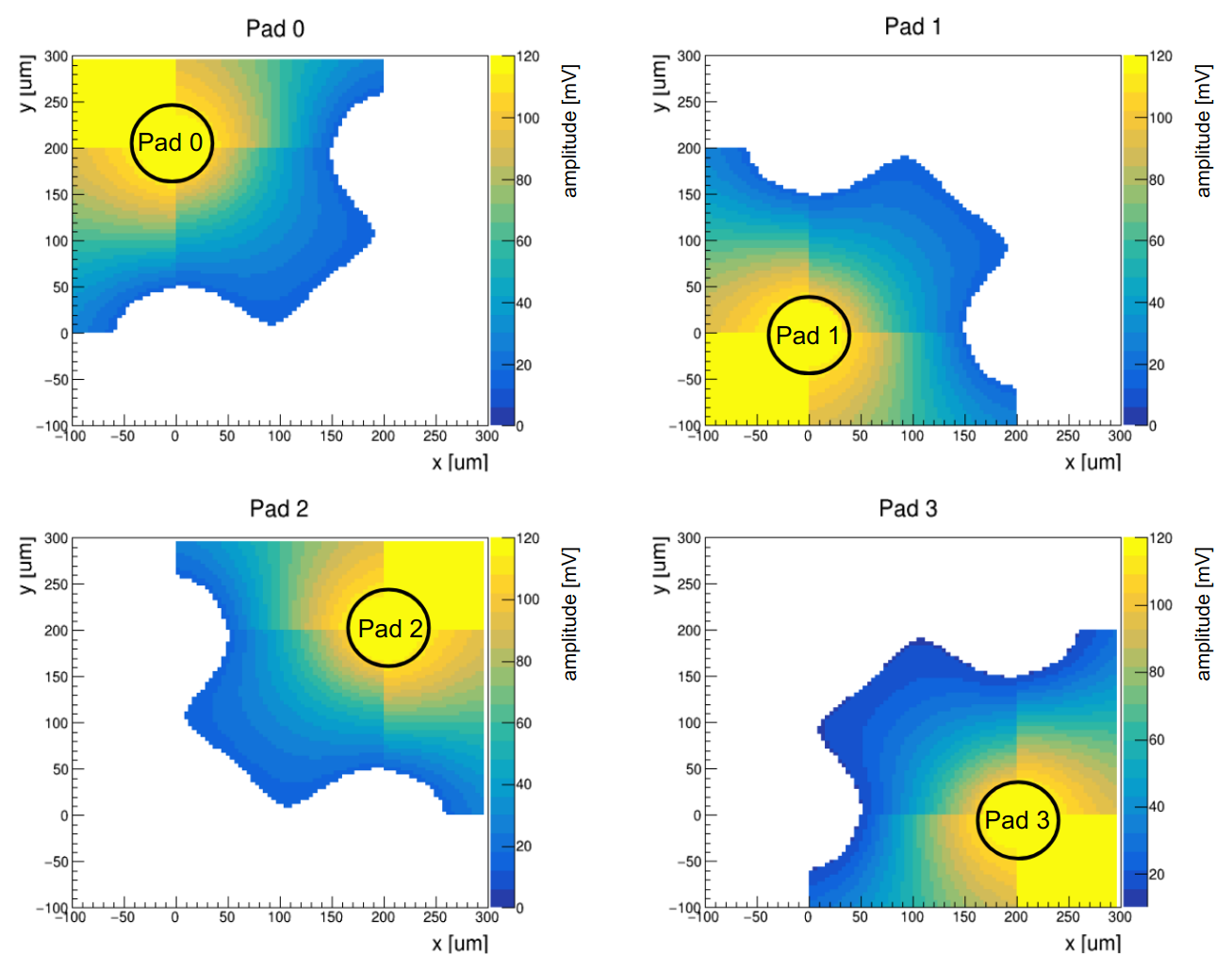}     
\caption{Amplitude seen by each of the 4 read-out pads of an RSD as a function of the impact position, considering a total signal amplitude of 120 mV. The colored axis represents the signal amplitude seen by the corresponding read-out pad when a particle hits the sensor in a given x-y position. The geometry of this sensor is 100-200 \mum~: the regions circled in black are the 100 \mum~ metal read-out pads, while the distance between the centers of two pads (the pitch) is 200 \mum. \label{fig:masterF_4pad}}
\end{center}
\end{figure}

Using this model, it is possible to compute the fraction of the signal seen on each pad for any RSD geometry. Figure \ref{fig:masterF_4pad} shows how a signal with an amplitude of 120 mV (corresponding to a gain of $\sim$15, which is a typical working point for LGAD devices) is shared in a 2x2 RSD sensor with 100-200 \mum~ geometry (circular pads have been drawn for simplicity, although the RSD1 prototypes have square pads). 

The second analytical model of signal sharing in RSD differs from the MF model by assuming  a linear attenuation of the signal with distance. This model is referred to as the "Linear Attenuation" (LA) model. 

To assure the selection of real signals, only amplitudes above 3-5$\sigma$ noise level should be used in the reconstruction. This minimum amplitude also determines the maximum distance from the impact point at which the pad can be used in the reconstruction.  In the present study, a threshold of 15 mV is used to safely reject the noise. The white areas in figure \ref{fig:masterF_4pad} indicates positions where the pad cannot be used in the reconstruction as the signal is below 15 mV. 

It is important to note that, according to Equation \eqref{eq:master}, when a particle hits directly a metal pad, the signal is visible only in that single pad as its resistance is null~\cite{cern-seminar}. As a consequence, the spatial resolution for particles hitting the metal is that  of a pixel detector with binary read-out: $\sigma_x \sim metal \; size / \sqrt{12}$. 

Since it is different between metal and non-metal regions, the RSD spatial resolution results to be dependent on the hit position.  In future productions, the design of the metal pads will be optimized in order to minimize such effect. 

Assuming a given signal total amplitude, an RSD analytical model is able to predict the amplitude on each pad near the impact point.  Using this information, the number of active pads (defined as those that see a signal above the minimum threshold) as a function of the hit position can be derived. 

\begin{figure}[h]
\begin{center}
\includegraphics[width=0.75\linewidth]{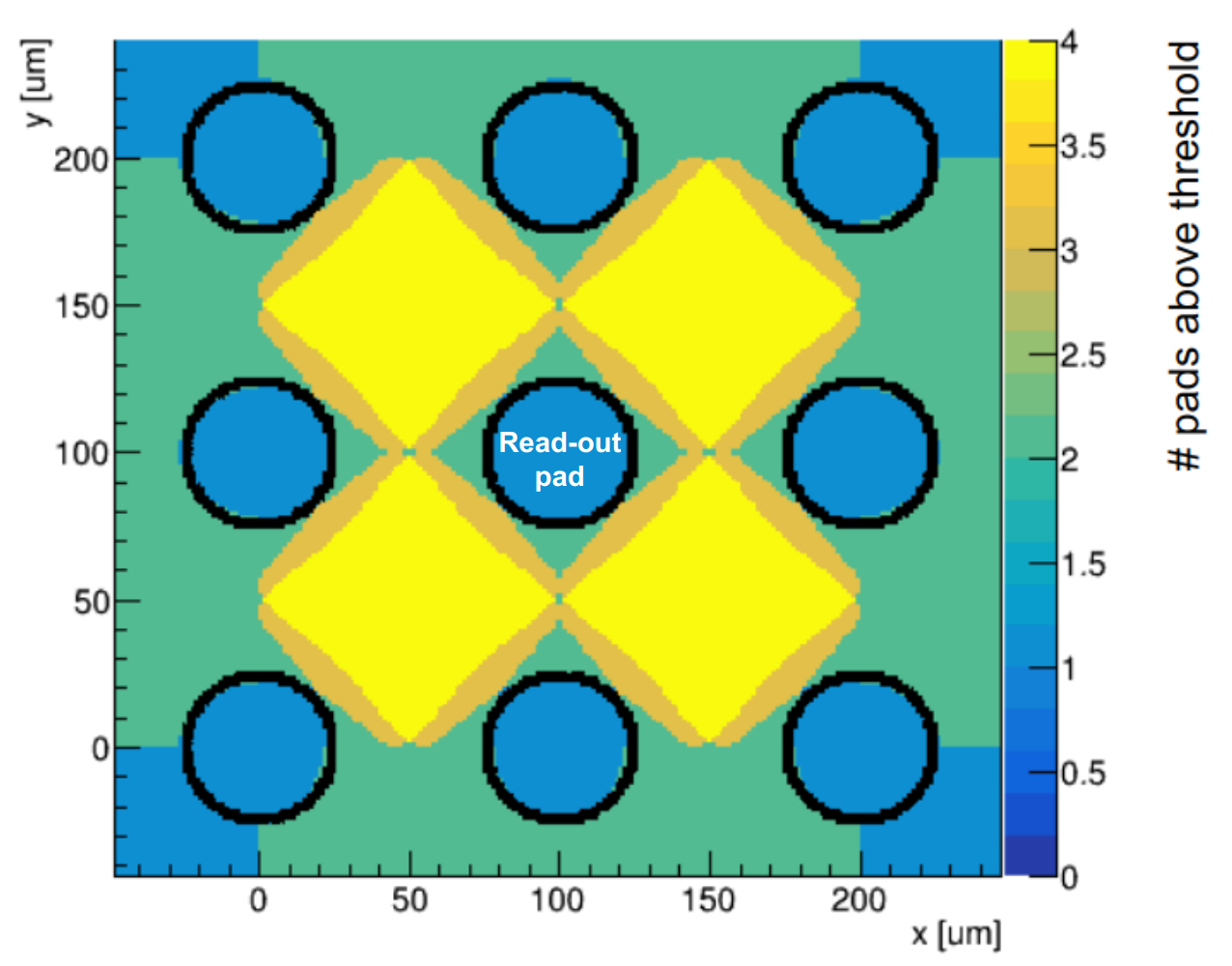}     
\caption{MF prediction of the number of pads that see a signal above 15 mV, assuming a total signal amplitude of 120mV (gain $\sim$15), in a 3x3 RSD with 50-100 \mum~ geometry. The colored axis represents the number of pads that see a signal larger than 15 mV when a particle hits the sensor in a given x-y position. The blue regions circled in black are the 50 \mum~ metal read-out pads: only one pad sees the signal there, because signal sharing does not occur underneath the metal. }
\label{fig:tappeto6}
\end{center}
\end{figure}

Figure \ref{fig:tappeto6} shows such 2D-map for a 3x3 pad RSD with 50-100 \mum~  geometry, considering a signal with a total amplitude of 120 mV and a threshold of 15 mV. Yellow and brown colors identify the efficient positions, where at least 3 pads see a signal; whereas in the blue and green parts only 1 or 2 pads are active, with consequent loss in position resolution.

These maps are useful to determine the regions where the position cannot be reconstructed unambiguously using the signal sharing technique (only 1 or 2 pads active) and to calculate the position reconstruction efficiency, defined as the fraction of active area not covered by metal with at least 3 active pads.

\begin{figure}[h]
\begin{center}
\includegraphics[width=0.85\linewidth]{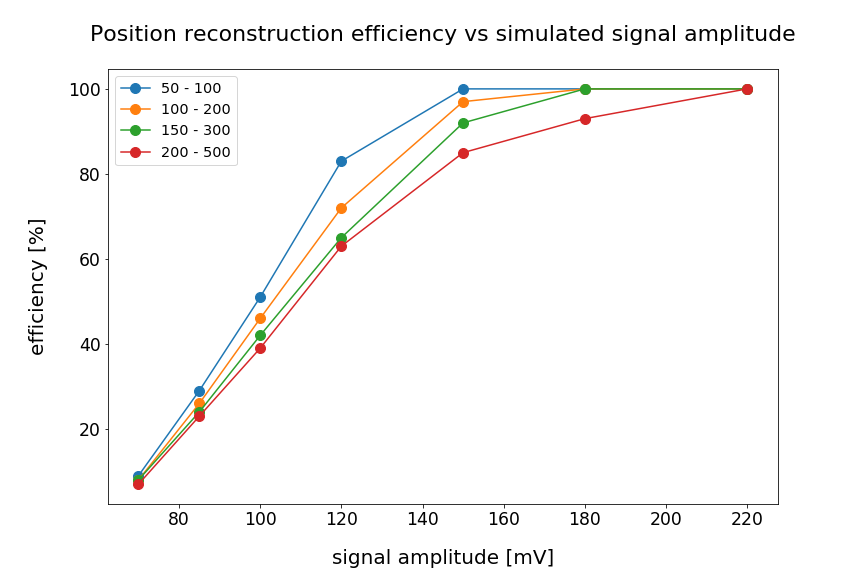}     
\caption{Position reconstruction efficiency as a function of the signal total amplitude for different RSD geometries.}
\label{fig:eff_vs_amp}
\end{center}
\end{figure}

A possible way to improve the performances of the reconstruction method is to increase the sensor gain: since the noise threshold is fixed to a given value, a larger signal will be above the noise threshold along a wider area. Figure \ref{fig:eff_vs_amp} shows the position reconstruction efficiency as a function of the signal total amplitude. The efficiency improves by increasing the sensor gain and devices with larger pitch require higher gain to reach full efficiency, as expected. Hence, a proper tuning of the gain is key to reach full efficiency. 

The gain definition in RSD is similar to that of standard LGADs \cite{ROPP}:

\begin{equation}
    Gain_{~RSD} = \frac{Q_{RSD}}{Q_{no-gain}}
\end{equation}

$Q_{RSD}$ is the total charge generated by a Minimum Ionizing Particle (MIP) and collected by the RSD at a given bias voltage; $Q_{no-gain}$ is the charge collected in the same conditions by an equivalent sensor which does not have the gain layer (the RSD1 production features several of these no-gain sensors specifically for the purpose of measuring the gain level). The RSD gain is therefore defined as the collected charge normalized to the charge collected by a sensor equivalent to the sensor under test, but without gain.

In this work, the devices under test were operated at 3 different gain levels: 12, 17 and 24, corresponding to a reconstruction efficiency ranging from 20-30\% for the lowest gain, to $\sim$ 100\% for the highest one. 

For the purpose of this work, focusing on the precision of the reconstruction algorithm, only the events falling in regions of the devices with 100\% 3-pads coverage are selected. 

Considering a sensor with given $pitch$ and the centers of its pads in: (0, 0) , (0, $pitch$), ($pitch$, 0) and ($pitch$, $pitch$), the regions used in the experimental measurements for each geometry are presented in table \ref{tab:ranges}.

\begin{table}[htbp]
\centering
\caption{\label{tab:ranges} X/y range of the central regions considered as experimental regions, assuming that the centers of the 4 read-out pads have coordinates: (0, 0) , (0, $pitch$), ($pitch$, 0) and ($pitch$, $pitch$).}
\smallskip
\begin{tabular}{c|c|c|c|c}
Geometry & $x_{min}$ [\mum] & $x_{max}$ [\mum] & $y_{min}$ [\mum] & $y_{max}$ [\mum] \\
\hline
50-100 \mum & 30  & 70 & 30 & 70\\
100-200 \mum & 60 & 140 & 60 & 140\\
150-300 \mum & 100 & 200 & 100 & 200\\
200-500 \mum & 150 & 350 & 150 & 350\\
\hline
\end{tabular}
\end{table}

This study will be repeated in the near future, as fully efficient RSD sensors will soon become available.

\section{Position reconstruction}
A detailed description of the RSD position reconstruction techniques is given in~\cite{marta}, here we summarize the principles. In RSD, a point on the surface is uniquely identified by the relative amplitudes seen by the nearby pads and this property can be exploited to reconstruct very accurately the impact point position. In the analytical models of reconstruction, the impact point is identified by comparing the measured signal sharing to a lookup table that contains for each position (for example every 1x1 $\mu m^2$) the predicted signal sharing. In~\cite{marta},  both MF and LA models were used to create  the look-up tables. 

The accuracy of the position reconstruction on RSD can be expressed as the sum of 3 terms:
\begin{equation}
    \sigma^2_{RSD}=\sigma^2_{jitter} + \sigma^2_{algorithm} + \sigma^2_{sensor}.
    \label{eq:total_res}
\end{equation}

\begin{itemize}
    \item $\sigma_{jitter}$: this term represents the spatial uncertainty due to the electronic noise. The effect of the electronic noise is to change the amplitude by $\sigma_{el-noise}$, creating an uncertainty in the localization that can be written as:
    
    \begin{equation}
   \sigma_{jitter} = \frac{\sigma_{el-noise}}{\frac{dV}{dx}}= \frac{\sigma_{el-noise}}{\frac{Amplitude}{Interpad}}
   \label{eq:jitter}
\end{equation}
     
    \item $\sigma_{algorithm}$: the reconstruction code uses algorithms to infer the hit position from the measured signals. This can be done analytically, for example using the MF lookup table, or with more advanced techniques such as those presented in this paper. In all cases, $\sigma_{algorithm}$ represents the uncertainty of the method selected;
    
    \item $\sigma_{sensors}$: this term groups all sensor imperfections contributing to an uneven signal sharing among pads.  The most obvious is a varying $n^{+}$ resistivity: a 2$\%$ difference in $n^+$ resistivity will turn an equal split between two pads into a non-equal split, yielding to an incorrect impact position reconstruction. 
\end{itemize}

In the following part of this paper, the $\sigma_{RSD}$ resulting from a reconstruction method based on a Multi-Output regressor is compared to that of the MF analytical method.

 \section{Position reconstruction using Multi-Output regression}
 \label{sec:ML}
In RSD, each pad sees a modified version of the original signal. During the propagation on the $n^+$ resistive surface, the signal becomes smaller, wider, with slower leading and falling edges, and is delayed. Each of these aspects is a valuable clue that should be used in the reconstruction of the hit position and time. For this reason, signal sharing among different pads leads naturally to the use of multivariate analysis techniques where many inputs contribute to the determination of the outputs.  

On the other hand, analytical methods such as MF and LA use only the information on the shared amplitude, and they cannot use extra parameters,  because the analytic description of how such parameters change with position is not very accurate. Thus, multivariate analysis techniques are intrinsically more powerful and can lead to better accuracy.  
 
In order to efficiently reconstruct the hit position, a Multi-Output Regressor algorithm, using a Gradient Boosting Regressor as an estimator, has been developed and trained ~\cite{sklearn}. 

Gradient Boosting is a regression technique, used to develop a prediction model, which works in a forward, stage-wise manner~\cite{gradient-boosting-1,gradient-boosting-2}. In a given analysis, N variables are selected to define an input vector $\Vec{x} = (x^{0}, x^{1}, x^{2},...,x^{N})$. A set of $n$ input vectors \{$\Vec{x}_{1},\Vec{x}_{2},\Vec{x}_{3},...,\Vec{x}_{n}$\} will produce $n$ output variables \{$y_{1},y_{2},y_{3},...,y_{n}$\}. The purpose of Gradient Boosting is to find a function $F(\Vec{x}_{i})$ such that:

\begin{equation}
F(\Vec{x}_{i}) = y_{i}, \;  i =1,...,n 
\label{eq:fxy}
\end{equation}

$y_{i}$ is always a scalar, whereas the input variables defining the input vector $\Vec{x}_{i}$ can either be one ($N = 1$) or more. 

In the following, the index $i$ will represent the $i-th$ element of the input or output dataset, while $k$ will indicate the $k-th$ component of a vector.

Gradient Boosting proceeds in forward stages, in order to increasingly smooth the difference between the output variables and the values predicted by $F$. This is achieved by introducing at each stage of the algorithm a \textit{weak learner} $h$, which represents the $residuals$, i.e. the discrepancy between $F(\Vec{x}_{i})$ and $y_{i}$ ~\cite{towards}:


\begin{equation}
    h(\Vec{x}_{i})=y_{i} - F(\Vec{x}_{i})
\end{equation}

The algorithm progressively reduces the $residuals$, in order to reach the best estimate for $F$.

The addition of $h(\Vec{x}_{i})$ to $F(\Vec{x}_{i})$ provides the functional form for the next stage of the algorithm. Therefore, at the $m-th$ step ($m$ represents the depth of the algorithm), $F$ is equal to:

\begin{equation}
    F_{m}(\Vec{x}_{i}) = F_{m-1}(\Vec{x}_{i})+h_{m}(\Vec{x}_{i}) = y_{i}
\end{equation}

The algorithm starts with an \textit{initial guess} $F_{0}$ (a constant value), which usually is simply the average value of the output variables $y_{i}$:

\begin{equation}
    F_{0}= \sum_{i} \frac{y_{i}}{n}
\end{equation}

Hence, $F_{m}(\Vec{x}_{i})$ can be written as:

\begin{equation}
    F_{m}(\Vec{x}_{i}) = F_{0}+\sum_{j=1}^{m}h_{j}(\Vec{x}_{i}) = y_{i}
\end{equation}

which clarifies the stage-wise nature of the algorithm. The sequential addition of \textit{weak learners} is helpful because it produces an ensemble of weak predictive models which is much stronger than the single model, thus resulting in better performances. 

Machine learning algorithms usually learn by means of a \textit{loss function}, which evaluates how well a model fits the training data. In this work, the \textit{loss function} is the \textit{least squares}:

\begin{equation}
    L(y_{i},F(\Vec{x}_{i})) = \frac{(y_{i}-F(\Vec{x}_{i}))^2}{2} 
\end{equation}

The sum over the number of elements in the input dataset of the \textit{loss function} $L$, is the ~~~~~~ function $J$:

\begin{equation}
    J = \sum_{i}L(y_{i},F(\Vec{x}_{i}))
\end{equation}

and it is straightforward to prove that:

\begin{equation}
     -\frac{\partial J}{\partial F_{m-1}(\Vec{x}_{i})} = y_{i} - F(\Vec{x}_{i}) = h_{m}(\Vec{x}_{i})
\end{equation}

Therefore, the $residuals$ can be treated, in this specific case, as the negative gradient of $J$. That is an important aspect, since, instead of  minimizing the $residuals$, the algorithm can aim at minimizing a gradient, which is easily accomplished by the Gradient Descent technique ~\cite{towards2}. Hence, the objective of Gradient Boosting is eventually the minimization, at each stage of the algorithm, of $\frac{\partial J}{\partial F(\Vec{x}_{i})}$.

In a typical regression problem, the output variable $y_{i}$ is a scalar. Multi-Output is an evolution of the standard regressor, as it is able to predict an output variable which is a vector of $N^\prime$ real-valued numbers: $\Vec{y}_{i}=(y_{i}^{(1)},y_{i}^{(2)},...,y_{i}^{(N^\prime)})$. 
In this work, multi-output regression has been implemented using the \textit{Single-Target Method} ~\cite{multi-output}. This method requires the multi-output models to be divided in $N^\prime$ single predictive models ($N^\prime$=2 in this work) which are trained on transformed training datasets $D^{k}$:

\begin{equation}
    D^{k}=\{(\Vec{x}_{1},y_{1}^{k}),...,(\Vec{x}_{n},y_{n}^{k})\}, k \in \{1,...,N^\prime\}
\end{equation}

Each model predicts a single output variable and, finally, all predictions are concatenated in order to provide the final, multi-output variable. The drawback of this approach is that the potential relations between the predicted variables are not exploited ~\cite{multi-output}. For the purpose of the present work, that aspect has not been found to be an issue.

\begin{figure}[h]
\begin{center}
\includegraphics[width=0.8\linewidth]{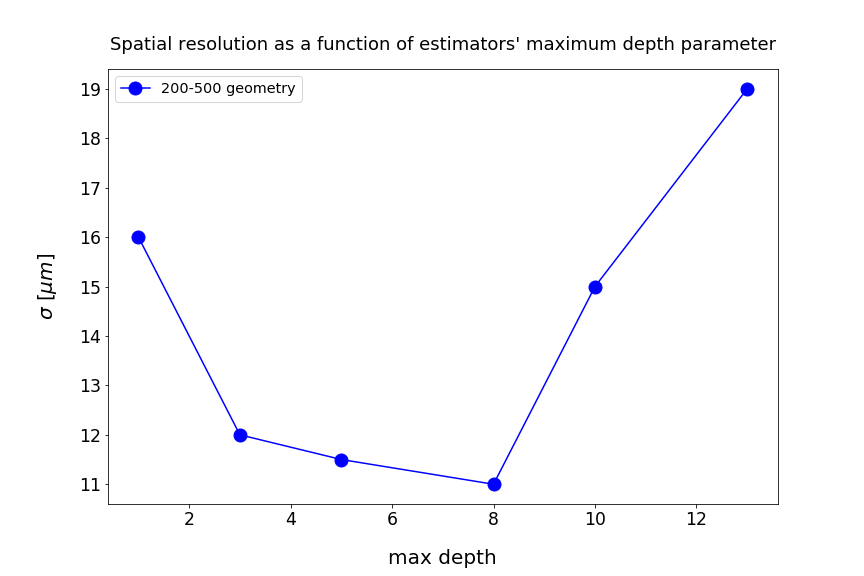} 
\caption{Spatial resolution as a function of the individual estimators maximum depth used in the Multi-Output Regressor model for the device with 200-500 geometry.  \label{fig:res_vs_depth}}
\end{center}
\end{figure}

The Multi-Output Regressor algorithm developed in this work takes 8 input variables: the 4 amplitudes seen by the read-out pads ($A_{l}$) and the same amplitudes normalized to the total amplitude ($A_{l}/\sum_{l=1}^{4} A_{l}$), while the x and y hit coordinates are the output. Such model has been trained with 100 boosting stages. The maximum depth (i.e. maximum number of stages) of the individual regression estimators was set to 8, with a learning rate of 0.1: such value provides the best resolution, as shown in figure \ref{fig:res_vs_depth}, whereas larger values would cause the model to overfit, without really improving the predictions (overfitting occurs when a machine learning algorithm generates data too similar or equal to the training dataset, leading to predictions which are not accurate \cite{ml1}).
 
While the final goal is to train the algorithm using precise data collected at beam test, in this initial proof-of-principle study, the algorithm was trained using a mixture of simulation and experimental data. 

The input signals for the algorithm training were obtained in the following way: 
\begin{itemize}
    \item The total signal amplitude (obtained by summing the amplitudes recorded by the 4 read-out pads) for each geometry and for 3 different gain levels was measured in laboratory using the laser setup described in the following section.
    \item  After that, 100 different signal amplitudes were generated for each position within the experimental regions reported in \ref{tab:ranges}; that was repeated for all geometries and for the 3 different gains.
    \item  Then, using the RSD Main Formula, the sharing among the pads was analytically computed.
    \item  Finally, a Gaussian noise smearing  was  added to each pad signal. The addition of noise, called \textit{training noise}, and the use of 100 different amplitudes per position are introduced to prevent overfitting. The \textit{training noise} prevents overfitting as it generalizes the machine learning model: its effect is comparable to training the algorithm with 100 different dataset, therefore it can also be seen as a form of data augmentation \cite{ml2,ml3}.
\end{itemize}
 
The input signals produced in this way, combined with their relative x-y coordinates, form the \textit{training dataset}, which is then used to train the Multi-Output Regressor algorithm.   

The \textit{training noise} is a key aspect because it is very helpful to make reliable predictions, but it also introduces an intrinsic uncertainty ($\sigma_{algorithm}$) since a set of 4 amplitudes does not uniquely define a position. Such uncertainty can be assessed by feeding to the trained algorithm the same set of amplitudes that were used for its training, this time without noise (this set will be called \textit{test dataset}, to distinguish it from the \textit{training dataset}) , and then comparing the predicted hit coordinates with the known coordinates. 



\begin{figure}[h]
\begin{center}
\includegraphics[width=\linewidth]{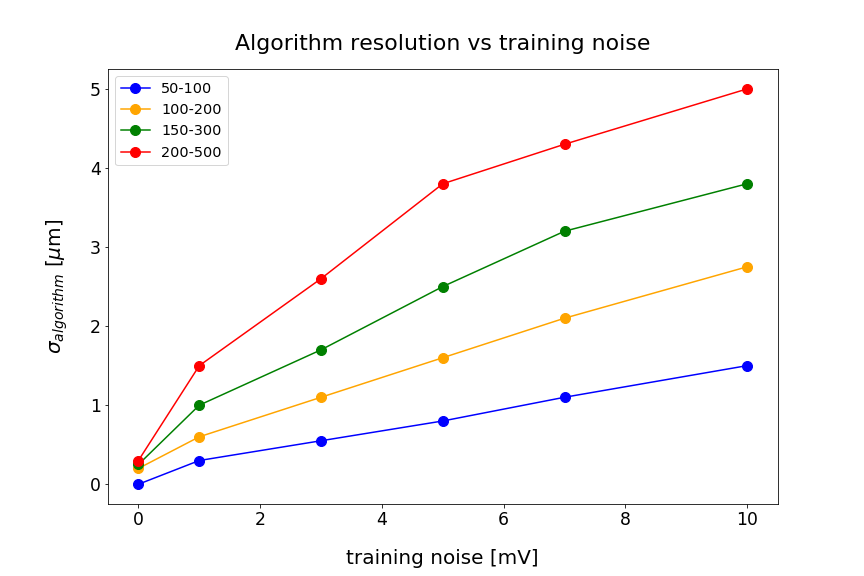} 
\caption{$\sigma_{algorithm}$ as a function of \textit{training noise} for RSDs with different designs. The signal amplitude is 120 mV for all geometries, corresponding to a gain $\sim$15.  \label{fig:intrinsic_vs_pitch}}
\end{center}
\end{figure}

Figure \ref{fig:intrinsic_vs_pitch} shows $\sigma_{algorithm}$ as function of the \textit{training noise} for the explored geometries, assuming a \textit{training dataset} with a total signal amplitude of 120 mV.  This analysis is performed using positions contained within a 40x40 \mum$^{2}$ square located in the middle of the area among the 4 pads. The term $\sigma_{algorithm}$ scales linearly with the \textit{training noise}, with larger geometries having a slightly worse resolution. Since the positions are within a squared region of fixed dimension for all geometries, such result does not depend on the width of the training region.

It is worth noticing that $\sigma_{algorithm}$ $\sim$ 0 for all geometries when \textit{training noise} equal to zero, underlining that, in absence of noise, the algorithm is able to predict the position with a negligible uncertainty, regardless the sensor's pitch. 

A possible explanation to the worse resolution in devices with larger pitch is the different attenuation of signal amplitude per unit of distance. A signal with total amplitude of 80 mV (gain 12) generated in a position equidistant from the centers of 4 different read-out pads will produce, according to equation \ref{eq:master}, a 20 mV signal amplitude in each of the pads, regardless the device geometry. In a RSD with the 50-100 \mum~ design, such hit position will be at a distance $d_{50-100}=25~\mu m*\sqrt{2}=35$ \mum~from the pad metal, whereas in a RSD with the 200-500 \mum~ geometry it will be $d_{200-500}=150 ~\mu m*\sqrt{2}=212$ \mum.  Assuming a linear change of the signal amplitude with distance, if the signal moves by one micron in the 50-100 (200-500) design, the signal will change by $\frac{1}{35} \sim 3 \% (\sim 0.5 \%$). Hence, a \textit{training noise} of 2 mV (a typical value used in this study) smears the reconstructed position  differently depending on the sensor size: 

 \begin{itemize}
    \item $\sigma^{50-100}_{algorithm}$: $\sim 2~mV / 0.6~mV/\mu m \sim 3.3~\mu m$
    \item $\sigma^{200-500}_{algorithm}$: $\sim 2~mV / 0.125~mV/\mu m \sim 20~\mu m$.
\end{itemize}

\begin{figure}[h]
\begin{center}
\includegraphics[width=\linewidth]{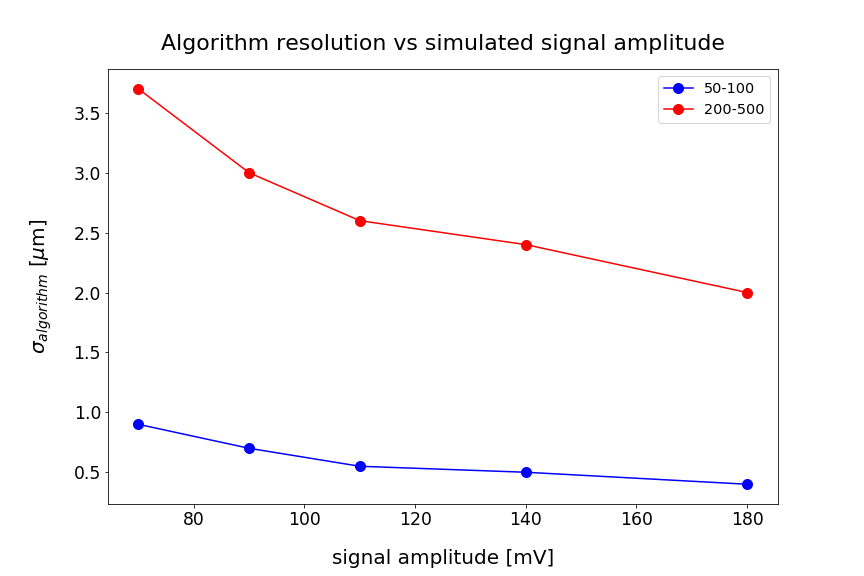} 
\caption{$\sigma_{algorithm}$ as a function of the simulated signal amplitude for RSDs having 50-100 and 200-500 \mum~geometries . The \textit{training noise} is set to 2 mV. \label{fig:intrinsic_vs_amp}}
\end{center}
\end{figure}

Another consequence of this effect is shown in Figure \ref{fig:intrinsic_vs_amp}: $\sigma_{algorithm}$ improves as a function of the simulated signal amplitude (i.e. the total amplitude used in the \textit{training dataset}) because the amplitude change per \mum~is higher and, as a consequence, the effect of the \textit{training noise} on the reconstructed position is milder. 

An equivalent smearing is also generated by the electronic noise: $\sigma_{jitter}$, see equation \ref{eq:jitter}, is higher for larger geometries as the \textit{interpad} increases.  

Smaller geometries are therefore benefiting twice: the reconstruction algorithm is more precise and the position jitter is smaller. In the same way, increasing the gain has a double effect: it decreases the contributions of the reconstruction method and of the jitter term.


\begin{figure}[h]
\begin{center}
\includegraphics[width=\linewidth]{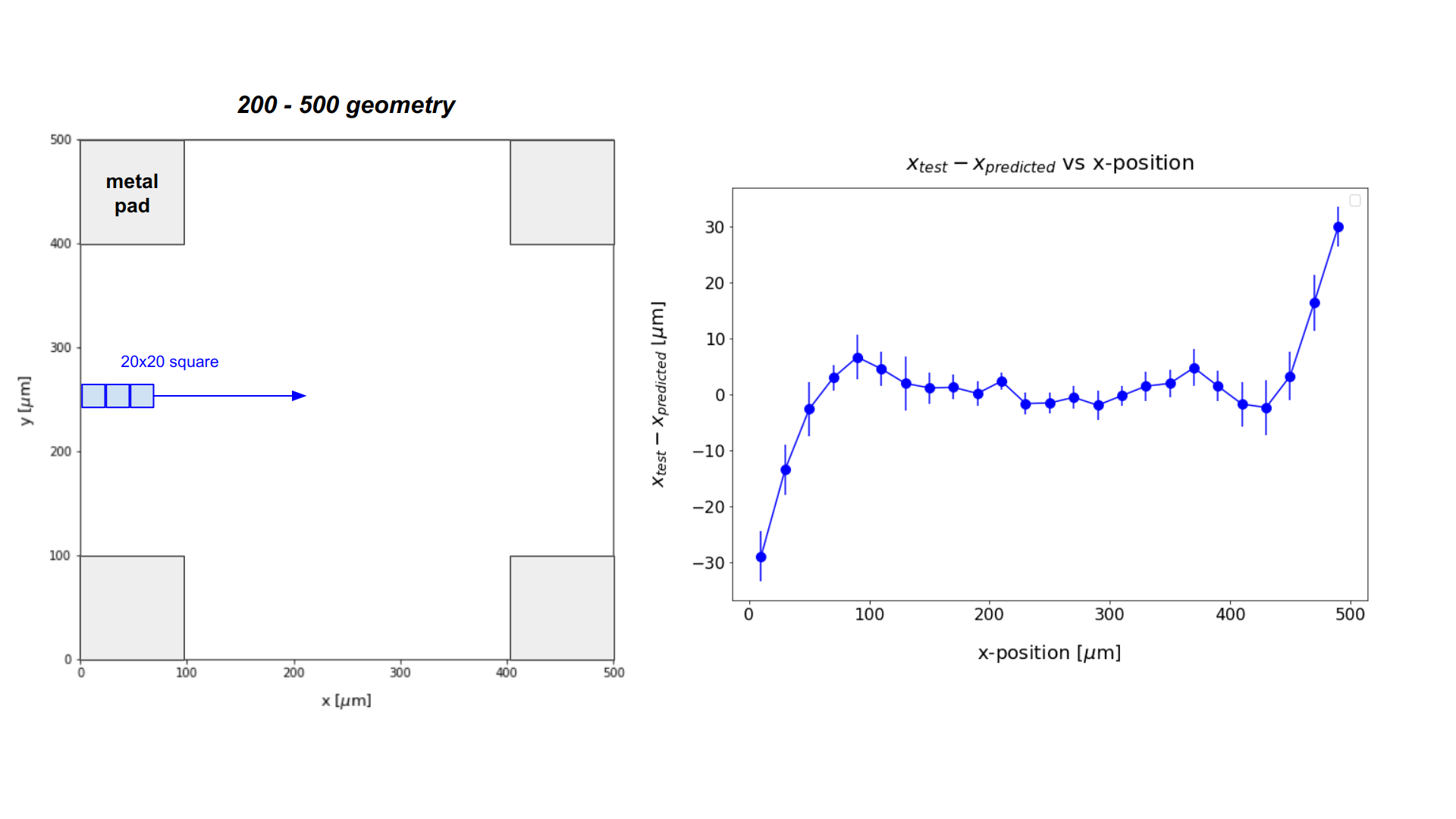}
\caption{Mean values of  $x_{test} - x_{predicted}$ distribution as a function of x-position for a 200-500 sensor. The simulated signal amplitude is 120 mV and the \textit{training noise} is 2 mV. }
\label{fig:offset}
\end{center}
\end{figure}

Up to now, $\sigma_{algorithm}$ has been computed using as test area  a 40x40 \mum$^{2}$ square positioned centrally among pads.
This study has been extended to the whole sensor area for the 200-500 geometry. 
The position resolution as a function of the x-axis has been computed in a sliding window of 20x20  \mum$^{2}$. This process is shown in the left side of figure \ref{fig:offset} while the right side of the picture shows, for each x position,  the offset and the value of $\sigma_{algorithm}$.  In the x-intervals 0-100 \mum~ and 400-500 \mum,  the offset worsens significantly since the signals in those areas are split among 6 pads while the test is performed considering only 4 pads. In the central area, 100-400 \mum,   $\sigma_{algorithm}$ is almost constant and the offset small.

\section{Results}

The performances of the Multi-Output regressor algorithm has been  measured using data collected with a laser setup. The events have been acquired shooting a 1064 nm picosecond laser on  several 3x3 RSD pixel detectors with different geometries (100-200 \mum, 50-100 \mum, 150-300 \mum, 200-500 \mum) and reading the signal of the 4 surrounding pads with a fast oscilloscope (4 GHz analog bandwith, 40 GS/s). The laser data are ideal for this measurement since they have been obtained using a Transient Current Setup (TCT), comprising a movable x-y stage, that measures the laser shot position ~\cite{gregor} with 2 \mum~ precision. It is therefore possible to use the positions provided by the TCT setup as reference positions to assess the accuracy of the reconstruction algorithm. The laser intensity was set to generate the same number of $e$-$h$ pairs of 3 Minimum Ionizing Particles (MIPs).
 
\begin{figure}[h]
\begin{center}
\includegraphics[width=\linewidth]{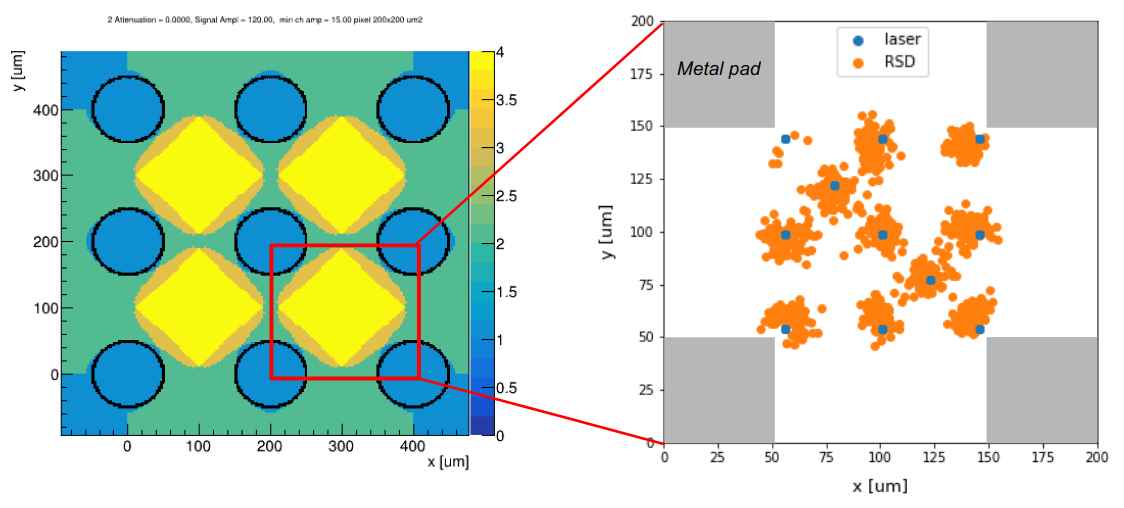} 
\caption{Left side: x-y map of the measured 3x3 RSD with 100-200 \mum geometry. The region considered during the measurement is highlighted in red. Right side: reconstructed RSD positions (orange) for different laser shots positions (blue). \label{fig:100-200_1}}
\end{center}
\end{figure}

Figure \ref{fig:100-200_1} (left side) shows the number of active pads as a function of x-y position for a 100-200 \mum~ RSD and the red square indicates the area tested with the laser. Similar regions have been considered in all measured sensors. 

In order for the algorithm to be fully efficient, the laser data have been collected only from positions within the experimental regions reported in table \ref{tab:ranges}, where at least 3 pads are active.

Figure \ref{fig:100-200_1} (right side) shows  the positions obtained with the ML algorithms  (orange) compared to the true laser shots positions (blue); a total of 100 laser shots were fired in each position. Similar maps have been obtained for the other geometries as well. As it is  clearly visible on the plot, the reconstructed points cluster tightly around the laser positions.

The width of the distribution $x_{RSD} - x_{Laser}$ ($\sigma_{measured}$) provides the squared sum of the RSD measured resolution ($\sigma_{RSD}$) and the TCT setup resolution ($\sigma_{TCT}$ = 2 \mum), hence:

\begin{equation}
    \sigma_{RSD} = \sqrt{\sigma_{measured}^{2}-\sigma_{TCT}^{2}}
\end{equation}
 
Table \ref{tab:resolutions} shows:  $\sigma_{RSD}$; the laser mean signal amplitude (used as total amplitude to generate the \textit{training dataset}); the \textit{training noise} used in the \textit{training dataset} that gave the best performance of the algorithm; the uncertainty of the ML algorithm ($\sigma_{algorithm}$) and the expected total resolution ($\sigma_{expected}$, details in the following section).  

\begin{table}[htbp]
\centering
\caption{\label{tab:resolutions} Spatial resolutions of RSD sensors.}
\smallskip
\begin{tabular}{c|c|c|c|c|c}
Geometry & $\sigma_{RSD}$ & Amplitude & \textit{Training} & $\sigma_{algorithm}$ & $\sigma_{expected}$ \\
&  ($\pm$ exp. $\pm$ syst.) &  &  \textit{noise} &  \\
\hline
50-100 \mum & 2 $\pm$ 0.1 $\pm$ 0.5 \mum & 190 mV & 1 mV & 0.2 \mum & 1.5 \mum\\
100-200 \mum & 4.5 $\pm$ 0.2 $\pm$ 1.5 \mum &  160 mV & 1 mV & 0.5 \mum & 3.5 \mum\\
150-300 \mum & 9.0 $\pm$ 0.4 $\pm$ 1 \mum & 225 mV  & 1 mV & 0.6 \mum & 3.5 \mum\\
200-500 \mum & 10.9 $\pm$ 0.2 $\pm$ 3 \mum & 135 mV & 30 mV & 27 \mum & 28 \mum\\
\hline
\end{tabular}
\end{table}

For the 50-100 device, $\sigma_{measured}$ is reported instead of $\sigma_{RSD}$, since the uncertainty is completely dominated by $\sigma_{TCT}$.

\begin{figure}[htb]
\begin{center}
\includegraphics[width=0.9\linewidth]{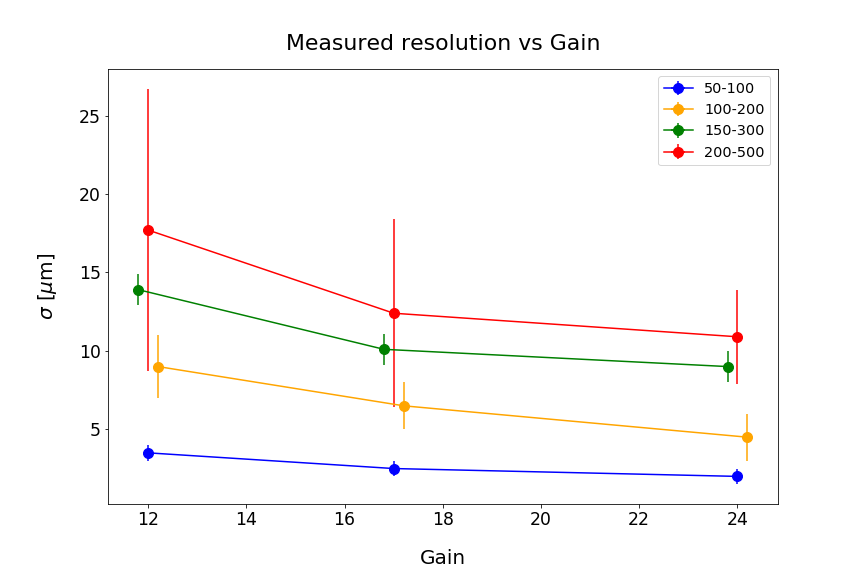} 
\caption{Measured spatial resolution for different geometries as a function of gain. \label{fig:res_vs_gain}}
\end{center}
\end{figure}

The behaviour of the spatial resolution as a function of gain is shown in figure \ref{fig:res_vs_gain} for all 4 geometries. As expected, the resolution improves with increasing gain since both $\sigma_{jitter}$ and $\sigma_{algorithm}$ decreases. 

\begin{figure}[h]
\begin{center}
\includegraphics[width=0.9\linewidth]{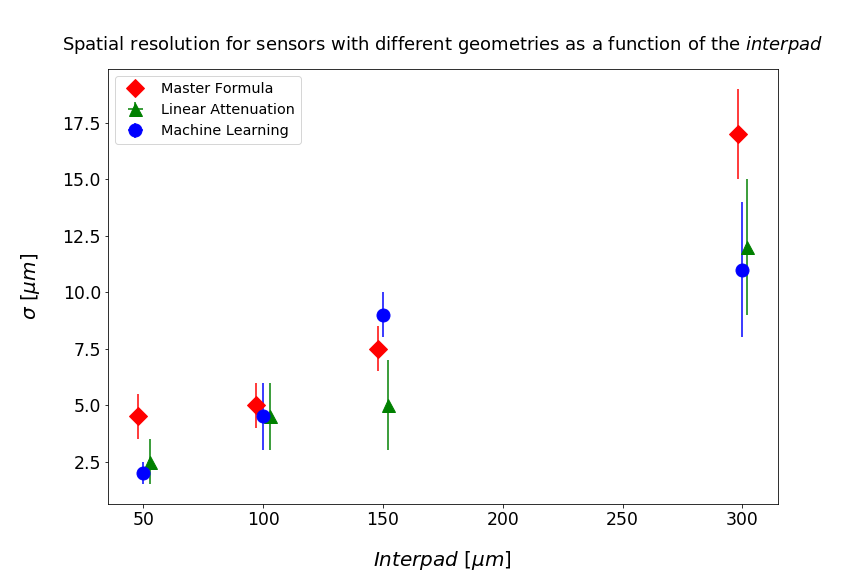} 
\caption{Spatial resolution as a function of the \textit{interpad} size for the measured RSD sensors. Red diamonds (green triangles) show the results obtained with the MF (LA) analytical method, while blue circles those obtained with ML.} \label{fig:res_vs_interpad}
\end{center}
\end{figure}

In \cite{marta}, two analytic methods (MF and LA) and the Discretize Positioning Circuit (DPC)  are used to evaluate the RSD position resolution.  Figure \ref{fig:res_vs_interpad} compares these results  with that of the ML method: the ML method is consistently the best or close to the best,  even though its training has not been performed on real data. This  demonstrates that an approach based on ML has the capability of improving the accuracy of RSDs beyond what is achievable with the more traditional reconstruction methods.  

\subsection{Discussion of the results}

\subsubsection{The optimum value of \textit{training noise}}
\label{subsec:training}
The results presented in Table~\ref{tab:resolutions}  were obtained with therein listed training noise.  The best performance of the algorithm is achieved when the \textit{training noise} is close to the experimental electronic noise, which was measured to be 1.5-2 mV. This result is expected, since the \textit{training noise} has on the \textit{training dataset} the same smearing effect that the electronic noise has on the experimental data, therefore it is not surprising that the algorithm predictions are most accurate when training and experimental noises are set equal or very similar. 


For the  geometry 200-500, on the contrary, the best resolution was reached  at a much higher \textit{training noise}, $\sim 30~ mV $. A possible explanation for this difference is the lower accuracy of the MF algorithm for the 200-500  geometry. As reported in figure \ref{fig:res_vs_interpad}, the use of the MF algorithm with the 200-500 leads to a large spatial resolution, indicating that this model fails to reproduce correctly the split among pads. 
Since the ML network has been trained with data generated using the MF model, also its resolution for the 200-500  geometry should be rather modest. However, the use of a large \textit{training noise} is able to compensate for this initial inaccuracy. This case highlights the power of the ML method: it achieves better results than a pure analytical model thanks to the noise introduced to the system. For the 
 200-500 sensor, the resolution achieved using the ML method is 55\% better than that with the MF.

\subsubsection{Systematic uncertainties}
The systematic uncertainties reported in table \ref{tab:resolutions} have been assessed by changing the \textit{training noise} in the vicinity of its optimal value and splitting each laser dataset in two subsets, and then comparing the results. Such uncertainties are rather small, demonstrating that the ML method is not very susceptible to simulation parameters. 

\subsection{On the expected spatial resolution}

The trained ML algorithm can be used not only to compute its intrinsic uncertainty ($\sigma_{algorithm}$), but also to give an expectation ($\sigma_{expected}$) on the spatial resolution that will be obtained once the algorithm is tested to real data. This can be done in the same way described in section \ref{sec:ML} for $\sigma_{algorithm}$, but this time using a \textit{test dataset} with a 1.5 mV smearing applied, which represents the electronic noise. 

While the smearing applied to the \textit{training dataset} (i.e. the \textit{training noise}) is responsible for $\sigma_{algorithm}$, the smearing applied to the \textit{test dataset} is comparable to the effect of the jitter, as described in equation \ref{eq:total_res}, therefore the expected resolution can be written as:

\begin{equation}
   \sigma_{expected} = \sqrt{\sigma_{algorithm}^{2}+\sigma_{jitter}^{2}} 
\end{equation}

Such equation does not take into account the third term of equation \ref{eq:total_res}, $\sigma_{sensor}$, which depends on non-uniformities of the real detector, that cannot be simulated. However, since $\sigma_{expected}$ is similar to $\sigma_{RSD}$ in the 50-100 and 100-200 designs, $\sigma_{sensor}$ is expected to be rather small in all geometries, because all tested sensors come from the same wafer and should therefore have similar non-uniformity terms.

$\sigma_{expected}$ is less accurate for 150-300, where it predicts a lower resolution, and 200-500, where conversely the expected resolution is much higher. The cause of such inaccuracies in the largest geometries should be sought in the imperfections of the MF, which might also be the reason for the large $\sigma_{algorithm}$ of the 200-500 device. Both aspects will be fixed in the future by training the algorithm directly with real data taken at the beam test.

\section{Conclusions and future plans}
A Multi-Output regressor algorithm has been trained to reconstruct with great accuracy the hit position in RSDs. A spatial resolution ranging between 2 \mum ~ and 10 \mum, depending upon the sensor geometry, has been measured. 

Even though this first implementation of the Multi-Output regressor algorithm uses a limited set of input variables in the training phase, it is presently the most accurate reconstruction algorithm for RSD. 

In the near future, the Multi-Output regressor algorithm will be trained with beam test data taken with a very precise tracker. In this way, the algorithm will be fed with a wider range of input features, experimentally measured, whose attenuation laws cannot be derived analytically. This will further improve the spatial resolution, especially in devices with large geometries, because it will be possible to train a much deeper and complex network.



\acknowledgments
We warmly thank the INFN Computing Center of Torino for providing support and computational resources and kindly acknowledge the following funding agencies and collaborations: INFN - Gruppo V; Horizon 2020, grant UFSD 669529; Dipartimenti di Eccellenza, University of Torino (ex L. 232/2016, art. 1, cc. 314, 337); Ministero della Ricerca, PRIN 2017, progetto PRIN2017L2XKTJ, 4DInSiDe. Ministero della Ricerca, FARE, progetto R165xr8frtfare.

\end{document}